\pdfoutput=1

\documentclass[12pt,preprint]{aastex}

\def\textfrac#1#2{{\textstyle\frac{#1}{#2}}}
\def\ainf{\alpha}
\def\acrit{\alpha_{\rm crit}}
\def\Dl{D_L} \def\Ds{D_S} \def\Dls{D_{LS}}
\def\hover#1{\setbox0\hbox to 0pt{\hss$\scriptscriptstyle
             \rightharpoonup$}%
             #1\kern0.4ex\raise 1.5ex\box0\kern-0.01ex}
\def\btheta{{\hover\theta}}
\def\kpc{\;{\rm kpc}}

\shorttitle{}

\shortauthors{}

\begin{document}

\title{The cluster lens ACO~1703: \\ redshift contrast and the inner profile}

\author{Prasenjit Saha}
\and
\author{Justin I. Read}
\affil{Institute for Theoretical Physics, University of Z\"urich, \\
       Winterthurerstrasse 190, 8057 Z\"urich, Switzerland}

\begin{abstract}

  ACO~1703 is a cluster recently found to have a variety of strongly
  lensed objects: there is a quintuply-imaged system
  at $z=0.888$ and several other
  lensed objects from $z=2.2$ to 3.0 (the cluster itself is at
  $z=0.28$).  It is not difficult to model the lens, as previous work
  has already done.  However, lens models are generically non-unique.
  We generate ensembles of models to explore the non-uniqueness.  When
  the full range of source redshifts is included, all models are close
  to $\rho\propto r^{-1}$ out to $200\kpc$.  But if the quint is
  omitted, both shallower and steeper models (e.g., $\rho\propto
  r^{-2}$) are possible.  The reason is that the redshift contrast
  between the quint and the other sources gives a good measurement of
  the enclosed mass at two different radii, thus providing a good
  estimate of the mass profile in between.  This result supports
  universal profiles and explains why single-model approaches can give
  conflicting results.  The mass map itself is elongated in the NW-SE
  direction, like the galaxy distribution.  An overdensity in both
  mass and light is also apparent to the SE, which suggests
  meso-structure.
\end{abstract}

\keywords{gravitational lensing --- galaxies: clusters: general}

\section{Introduction}

An important prediction of hierarchical clustering of cold dark matter
is a universal profile for halos. Universal profiles were originally
indicated by cosmological simulations
\citep{1997ApJ...490..493N,1998ApJ...499L...5M} but more recently
phenomenological models have also produced them
\citep[e.g.,][]{2005ApJ...634..756A,2006MNRAS.368.1931L}, as have more
specialized simulations \citep{2006ApJ...653...43M}.  The precise form
of a universal profile and the expected variation from halo to halo
are still being debated; \cite{2006AJ....132.2685M} discuss some
competing parametrizations.  A further complication arises in that
infalling baryons would steepen the dark-matter profile somewhat
(adiabatic contraction), but in large clusters this effect is expected
to be small exterior to $\sim 20\kpc$ \citep{2004ApJ...616...16G}.
There is general agreement that in the inner regions of clusters the
dark matter profile would have $\rho\sim r^{\rm -1\ to\ -1.4}$
\citep{2004MNRAS.353..624D}.

Lensing is an obvious way to measure the profiles of clusters, and in
this paper we will address specifically multiple-image (i.e.,
``strong'') lensing.  Weak lensing by outer regions of clusters is
also of interest \citep[for a recent example,
see][]{2008MNRAS.385.1431H}.  Other possibilities are to use
generalized virial theorems and the kinematics of cluster galaxies
\citep{2006MNRAS.367.1463L}, or hydrostatic equilibrium of x-ray gas
\citep{2008MNRAS.386.1092L}.  Combinations of these approaches are
also possible.

Extracting a mass profile from lensing is not trivial, however.  The
reason is that lensing depends on {\em integrated\/} mass.  For
example, the light deflection near the Sun
$4GM_\odot/(c^2R_\odot)=1.75''$ says nothing about the mass profile of
the Sun.  Similarly, a single Einstein ring measures only the mass
enclosed within a cylinder.  Having two or more images at different
radii provides some more information, but is still not sufficient to
specify the mass profile, because it is possible to redistribute the
mass in a way that leaves the image configuration invariant (steepness
degeneracy).  Much more information becomes available if the
background sources are at different distances behind the lens.  Light
from further sources experiences a larger
effective lensing deflection and hence a larger Einstein radius.
Multiple Einstein radii, resulting from multiple source redshifts or
``redshift contrast'', measure the mass within multiple cylinders,
thus providing a robust constraint on the mass profile.  ACO~1703
turns out to be a striking example of this type of constraint.

The non-uniqueness of mass models that fit a given set of observables
has been much studied in lensing theory.  The steepness degeneracy was
discovered by \cite{1985ApJ...289L...1F}, and then rediscovered
multiple times; \cite{2000AJ....120.1654S} gives a short history of
these (re)discoveries, and a unified picture of degeneracies then
known.  Further degeneracies continue to be uncovered from theory
\citep{2008MNRAS.386..307L}.

But in practice, lensing degeneracies are most likely to be noticed in
the modeling process, when different kinds of models are tried out.
\cite{1991ApJ...373..354K} was the first to try a range of models and
distinguish robust features from model-dependent features.  Subsequent
work found parameter degeneracies appearing in models in many
different ways
\citep{1994AJ....108.1156W,1999AJ....118...14B,2000ApJ...535..671T}.
Recently \cite{2008ApJ...674..711S} report non-uniqueness in models of
clusters.

Faced with the non-uniqueness of models, three strategies are possible.
\begin{enumerate}
\item One could justify the model-type being fit from astrophysical
  arguments, and then assume it is correct.  This approach goes back
  to the first models of the first two known lenses
  \citep{1981ApJ...244..723Y,1981ApJ...244..736Y}. With the benefit of
  hindsight, we find that these pioneering models inferred correctly
  that the ``triple-quasar'' was really a quad, but predicted time
  delays for the double quasar several times too long.
\item One could try and introduce new information such as stellar
  kinematics or x-rays, as indeed \cite{1981ApJ...244..736Y}
  advocated, to reduce model-dependence in the results.
  \cite{2008MNRAS.384..987C} is a recent example.
\item One can explore model degeneracies and quantify uncertainties by
  generating large ensembles of models, thereby also identifying the
  best-constrained systems.
\end{enumerate}

With this background in mind, the recent discovery by
\cite{2008arXiv0802.4292L} of 13 different multiple-image systems in
ACO~1703 is very exciting.  Though not a well-known cluster, ACO~1703
is a strong x-ray source \citep{1992MNRAS.259...67A} and hence a
natural place to search for lensing. For modeling, the authors adopted
strategy~1 above, and their model is a generalized NFW profile, having
\begin{equation}
\rho = \frac{\rho_0}{{\left(\frac{r}{r_s}\right)^\alpha}\left(1+\frac{r}{r_s}\right)^{3-\alpha}}
\label{eq-nfwlike}
\end{equation}
where $r$ is an elliptical radius. The normalization $\rho_0$ is
defined indirectly in terms of the concentration $c$, such that the
average density within $r=cr_s$ is 200 times the
critical cosmological density.  The models give $\alpha\simeq1.1$,
$r_s\simeq170''$ or $\simeq 700\kpc$, and $c\simeq3$.  These results
appear to verify the prediction of a universal profile, but they also
raise some questions: (i)~Is the model an acceptable fit to the data,
according to a $\chi^2$ or other goodness-of-fit test?  (ii)~Can
other, very different, models provide equally good fits?  (iii)~What
do the fitted values of $r_s$ and $c$ really mean, when the data go
out to only $\sim50''$ ($\sim200\kpc$)?

In this paper we analyze the same multiple-image data following
strategy~3.  The technique is basically the same as in our previous
work on the inner profiles of J1004+411 and ACO~1689
\citep{2006ApJ...652L...5S}. But in this work we study more carefully
just where the constraints come from.

\section{Modeling the cluster}

\subsection{The image configurations}

Of the 13 multiply-imaged background objects, making a total of 42
lensed images identified by \cite{2008arXiv0802.4292L}, some sources
are probably groups of galaxies, leading to similar image systems.  Of
completely independent sources, there are at least six.  Their image
configurations are typical of cluster lenses, and can be summarized as
follows.

\begin{itemize}

\item {\em The quint.} A five-image system \citep[1 in the numbering
  system of][]{2008arXiv0802.4292L} with a spectroscopic redshift $z=0.888$.
  The morphology is like the well-studied galaxy lens J0911+055,
  except that the fifth image (which in galaxy lenses is almost never
  observable) is here distinctly present.

\item {\em The central quads.} Two four-image systems (15, 16),
  resulting from a two-component source.  This configuration is common
  in galaxy lenses.

\item {\em The incipient quint.}  One two-image system (2), probably
  two-nearly merging images of a quint. The well-known galaxy lens
  PG1115+080 would look like this, if we knew only the two brightest
  images in that lens.

\item {\em The northern lemniscates.} Three three-image systems (7, 8,
  9), again from a multi-component source, with a typical naked-cusp
  or lemniscate configuration.  This type of image configuration
  requires a strong quadrupole in the lens potential and is not shown
  by galaxy lenses, but is common in cluster lenses.

\item {\em The southern lemniscates.} Two three-image systems (10, 11)
  with morphology similar to the northern lemniscates.

\item {\em The northern long arcs.} Two three-image system (4, 5)
  which could also be lemniscates, but where modelling suggests
  further incipient images.

\end{itemize}

For two systems (nos.~3 and 6) in \cite{2008arXiv0802.4292L}, the
image configurations (that is, image parities, time ordering, etc.)
are less clear, though both are somewhat similar
to the northern lemniscates. For this reason, we do not consider these
two image systems in this paper.  However, model ensembles that do
include these systems, with plausible image parities and time
ordering, agree with the results presented here within quoted
uncertainties.  Many more candidate lensed objects are evident in the
cluster image, and it seems likely that several more lensed systems in
this cluster will be identified in the future.

\begin{figure}
\begin{center}
\epsscale{0.2}
\plotone{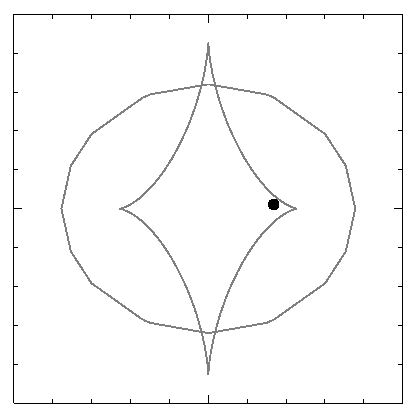}\plotone{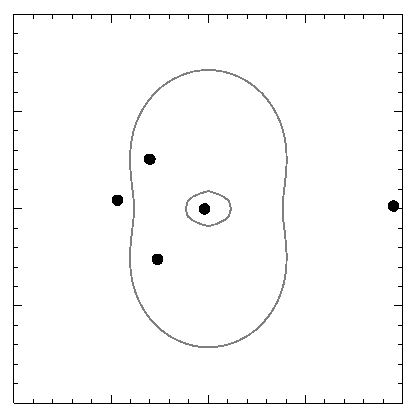}\plotone{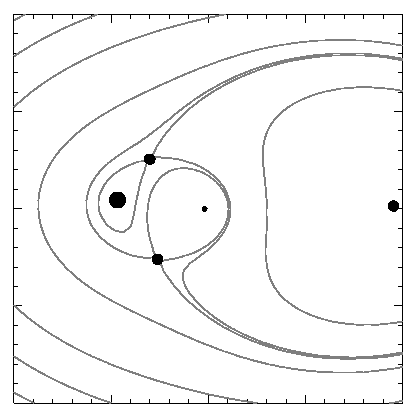}
\goodbreak
\plotone{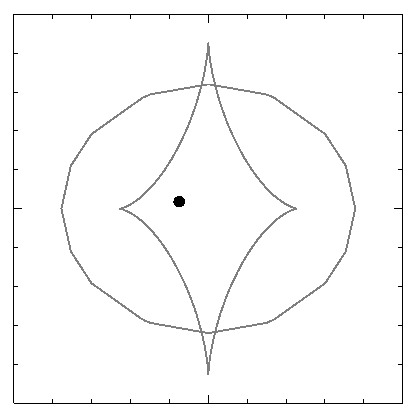}\plotone{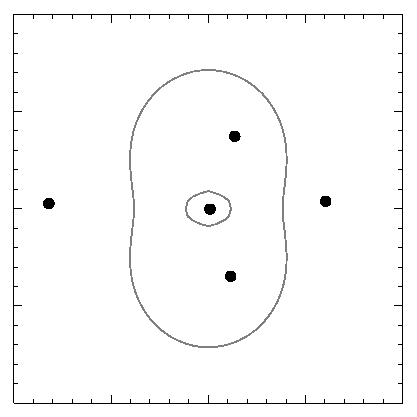}\plotone{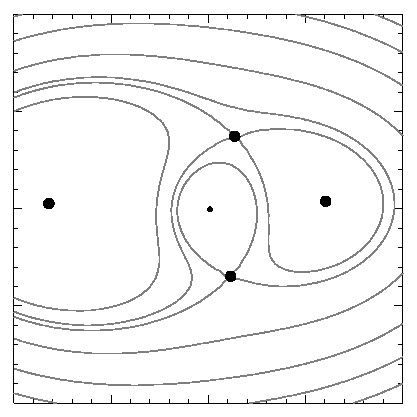}
\goodbreak
\plotone{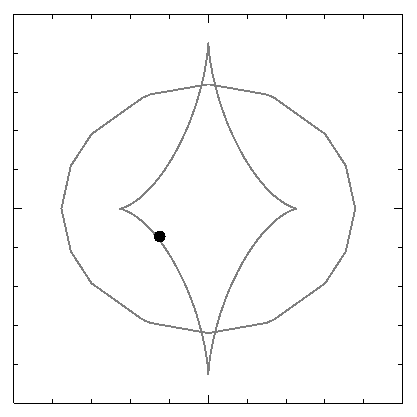}\plotone{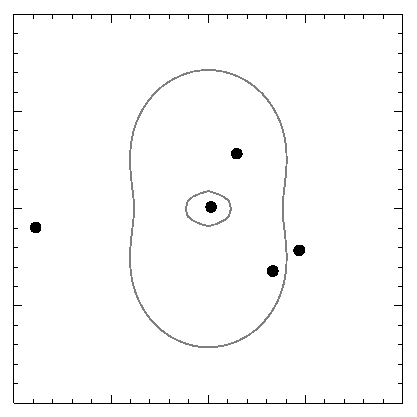}\plotone{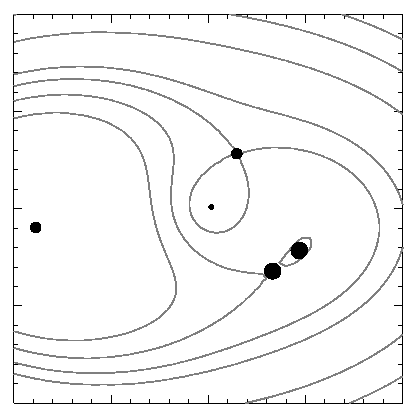}
\goodbreak
\plotone{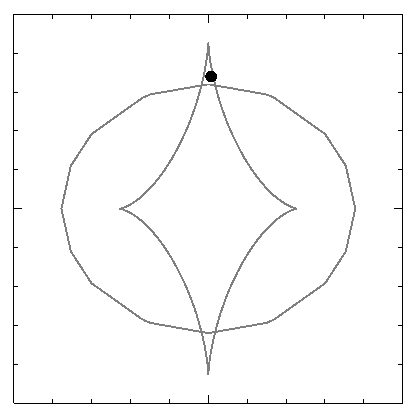}\plotone{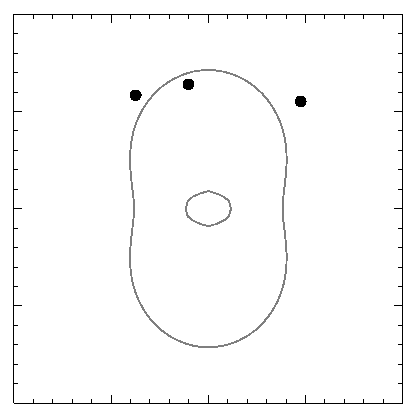}\plotone{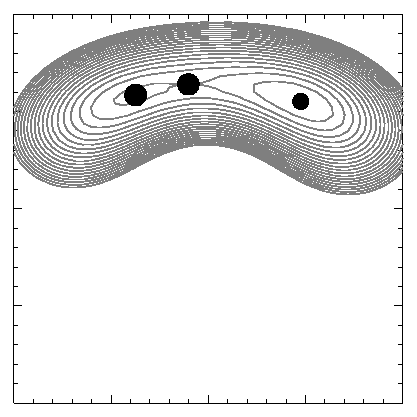}
\caption{Qualitative summary of the image systems, using a very simple
lens model.
\endgraf
The model is a cored isothermal sphere with external
shear, the lens potential being
$$ \psi(\btheta) = \left(\theta_x^2 + \theta_y^2 + \theta_c^2\right)
                   ^{\frac12} +
                   \textfrac12\gamma(\theta_x^2-\theta_y^2)
$$ with $\theta_c=0.1$ and $\gamma=0.3$.
\endgraf
The top row of panels corresponds to the quint, followed by the quad,
the incipient quint, and the northern lemniscate.  (The southern
lemniscate and the northern long arc are qualitatively similar to the
northern lemniscate.)  In each row, the left panel shows the source
position and the caustics, the middle panel shows the image positions
and the critical curves, while the right panel shows the image
positions and arrival-time contours.  The scales are not all the same,
but in all panels the distance between ticks is 0.1 in model units.}
\label{fig-simp}
\end{center}
\end{figure}

Fig.~\ref{fig-simp} conveniently summarizes the image configurations,
with the help of a very simple model: a cored isothermal sphere with
external shear.  Such a model can roughly reproduce the image
positions, except that the relative sizes of the image systems are
incorrect.  In particular, Fig.~\ref{fig-simp} gives roughly the same
size for the quint and the quad, whereas the real quint is much
smaller than the quad.  The reason is that the source redshift is much
smaller for the quint than the quad, and this fact becomes significant
below.

\subsection{The modeling method}

We model the cluster using the {\em PixeLens\/} method.  The technique
is described in detail in \cite{2008ApJ...679...17C}, and involves two
ideas.  (Neither of these is unique to {\em PixeLens,} though the
combination is.)

\begin{enumerate}
\item Instead of being cast as a parameter-fitting problem, lens
  reconstruction is formulated as an inversion problem.  The image
  data are treated as constraints on the mass distribution.  For
  example, the five images of the quint are required to map to a
  common source position.  Astrometric errors are assumed negligible.
  (Hence the goodness-of-fit problem does not appear, as the mass map
  is required to fit the image data precisely.)  In addition, a mass
  map is required to satisfy some prior constraints.  Specifically, in
  this work we require that (i)~the mass distribution must be
  non-negative everywhere, (ii)~the local density gradient must
  point within $60^\circ$ of the direction to the brightest cluster
  galaxy, and (iii)~no pixel other than the central pixel can be
  higher than twice the sum of its neighbors.
\item Rather than a single model or a few models, an ensemble of 100
  models is generated, which automatically explores degeneracies and
  provides uncertainties.  If a single model is desired for
  illustration, the ensemble-average model can be used; this model is
  in Bayesian terms the expectation over the posterior.
\end{enumerate}

Formulation of mass reconstruction in strong lensing as an inversion
problem was
introduced by \cite{1997MNRAS.292..148S} and extended to combined
strong and weak lensing
\citep{1998AJ....116.1541A,2001ASPC..237..279S}.  These works used a
form of regularization to find a single model consistent with the data
and optimal according to some criterion (for example, minimal
variation of $M/L$).  Varying the regularization gave some idea of the
uncertainties.  The basic scheme has been adapted in different ways,
\citep[e.g.,][]{2005A&A...437...39B,2008arXiv0802.0004D,2008arXiv0803.1199C}.
An interesting new idea comes from \cite{2007MNRAS.380.1729L} who
incorporate a constraint that additional bright images are not
produced.

The technique of generating ensembles of models to explore
degeneracies and estimate uncertainties (no regularization is
involved) was introduced in \cite{2000AJ....119..439W} and later
packaged in {\em PixeLens\/} \citep{2004AJ....127.2604S}, although a
theoretical justification for the precise model-sampling strategy was
not available until \cite{2008ApJ...679...17C}.  Other kinds of model
ensembles, involving parametric models, have also been used
\citep{2003ApJ...590...39K,2004ApJ...605...78O}.

\subsection{Model ensembles for ACO~1703}

\begin{figure}
\begin{center}
\includegraphics[width=0.49\textwidth]{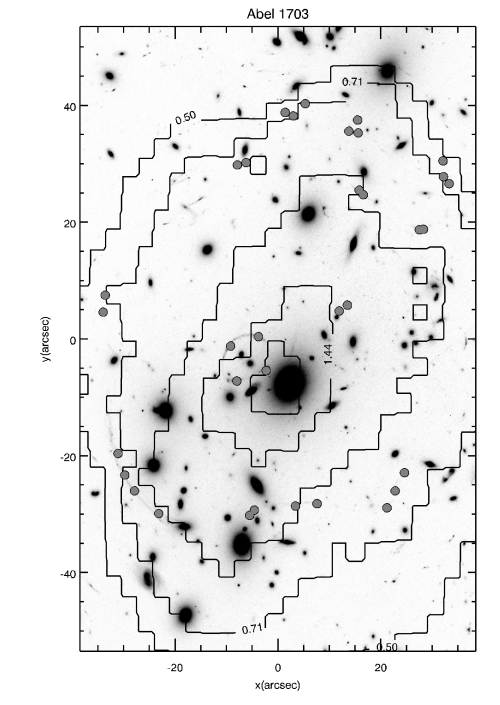}
\includegraphics[width=0.49\textwidth]{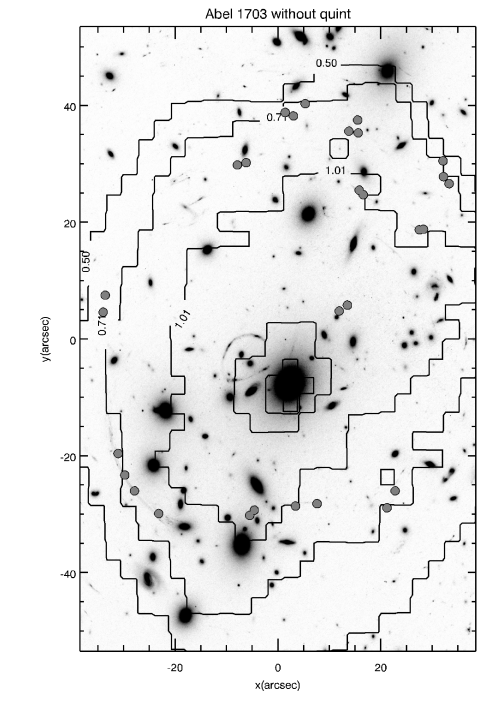}

\caption{Mass reconstruction of ACO~1703, superimposed on an optical
  image of the cluster \citep{Stott:2007}.  North is up and East to the left.
  The two panels correspond
  to all image systems included (left panel) and quint excluded (right
  panel).  Gray filled circles show image positions.  Contours refer
  to the ensemble-average PixeLens mass model.  These are labelled in
  units of the critical density for sources at infinity, or
  $3.4\times10^{10}M_\odot/\rm arcsec^2$.  The distance scale is
  $4.2\kpc/\rm arcsec$.  Note the overdensities in both mass and
  light, especially to the SE.}

\label{fig-mass}
\end{center}
\end{figure}

We now model ACO~1703 from the lensing data, considering two cases in
detail: first with all the image systems discussed above included, and
then with the quint excluded.  The possible models in the `with-quint'
case are naturally a subset of the possibilities in the `no-quint'
case.  Figs.~\ref{fig-mass} to \ref{fig-crit} show some results from
the ensemble-average models.

Fig.~\ref{fig-mass} shows the ensemble-average mass maps for the two
cases.  Fig.~\ref{fig-arriv} shows the morphologies in the {\em
PixeLens\/} models; the with-quint case is shown, but the no-quint
case is similar (apart from the quint itself).  Then Fig.~\ref{fig-crit}
compares the critical curves for the quint and one of the
quads---these are the first and second objects shown in
Fig.~\ref{fig-arriv}.

Qualitatively, we see that the cluster is similar to the simple model
illustrated in Fig.~\ref{fig-simp}, provided we take 11~o'clock in
Fig.~\ref{fig-simp} as `north'.  The arrival-time contours for the
{\em PixeLens\/} models are like those of the simple model, as are the
image configurations.  The long axis of the lens is oriented
perpendicular to the long direction of the quint in both cases.  The
critical curves are also similar, although the {\em PixeLens\/} curves
are jagged because of pixelation.  (The arrival-time contours are not
jagged because they depend on integrals over the mass distribution and
hence are smoother.)  Clearly, these qualitative features depend only
on the image identification and would appear in all models.

But of course, examining the {\em PixeLens\/} models in detail reveals
many more features.  First, we see in Fig.~\ref{fig-mass} that the
with-quint and no-quint models look similar but the latter look
slightly shallower.  This is only true of the ensemble average; as we
will see below, the no-quint ensemble contains a much larger variation
of models, including steeper and shallower models than the with-quint
ensemble.  A second feature is that in Fig.~\ref{fig-mass} the mass
contours seem to trace out the distribution of galaxies.  (Recall that no
information about the cluster galaxies, except the location of the
brightest galaxy, was used in the modeling.)  This suggests
meso-structure, or extended dark-matter structure correlated with
galaxies, similar to J1004+411 and ACO~1689
\citep{2007ApJ...663...29S}.  We will not attempt to analyze this
possible meso-structure in the present paper, but we will see some
further indications below.  A third feature is that the quint is
smaller than other image systems.  In Fig.~\ref{fig-arriv} the quint
is shown zoomed to make it comparable to the others in size, while in
Fig.~\ref{fig-crit} we see that the quint has smaller critical curves.
This is all simply because the quint has a smaller source redshift,
and hence a smaller $\Dls/\Ds$, resulting in a smaller Einstein
radius.  Thus, as a consequence of what we may call the redshift
contrast between the quint and other sources, we have distinct
Einstein radii within which the enclosed mass is well-measured.  This
enables a good estimate of the steepness of the mass profile, as we
see below.

\section{Source-redshift contrast and the inner slope}

We now specialize to the radial profiles, but unlike the previous
section we take the ensemble-derived uncertainties into account.

\subsection{Deprojecting the mass map}

From the mass map we now take the circular average to obtain
$\Sigma(R)$, and then deproject to derive $\rho(r)$ by numerically
solving the usual Abel integral equation. Appendix~\ref{ap-tests}
gives details and tests of the deprojection method.

Fig.~\ref{fig-surfden} shows $\Sigma(R)$ and $\rho(r)$ for ACO~1703
with 68\% and 99\% confidence intervals, when the cluster is
reconstructed with and without the quint.  Notice what a difference
the quint makes: without the quint, the mass profile between the
innermost and outermost images could be like $r^{-1}$, but it could
also be like $r^{-2}$ within the uncertainties; when the quint
is included, $\rho(r)$ is inferred as close to $\propto r^{-1}$.

Note that to arrive at the conclusion that the redshift constrast
reduces the available `model-space', the model-ensemble strategy is
essential.  Within the paradigm of a single best-fit model, one cannot
formulate such a conclusion.  However, a single-model strategy may
still provide a hint; \cite{2008arXiv0802.4292L} found that the slope
of their best-fit model was sensitive to the redshift of the quint,
and this seems to be such a hint. We emphasise that redshift contrast
does not act simply by increasing the radius range.  Disregarding the
three innermost images of the quint gives a radius range almost
identical to the no-quint case, but still adds a redshift contrast. In
this case the uncertainties within the image region still shrink
dramatically, as with the full quint data.

The spherically averaged density profile $\rho(r)$ in
Fig.~\ref{fig-surfden} also shows a `shelf' (a shallowing of the
log-slope) for $r \sim 100\kpc$.
This radius corresponds to the possible meso-structure, which we have
suggested above is indicated by Fig.~\ref{fig-mass}.  Tests
(cf.~Appendix~\ref{ap-tests}) show that deprojection of substructure
will lead to a shelf or even a bump in $\rho(r)$.

It is possible to fit the spherically averaged density using the
NFW-like form (\ref{eq-nfwlike}), and values like $\alpha=1$,
$r_s=500\kpc$, $c=4$ \cite[such as][derive]{2008arXiv0802.4292L}
are plausible within the uncertainties.
Formal least-squares parameter fits can be carried out, but are not
always meaningful, because they involve the tacit assumptions that the
radial bins are uncorrelated, and standard deviation over the model
ensemble represent Gaussian dispersions, both of which are poor
approximations.  With this caveat in mind, the
formal best fit to the region between the innermost and the outermost
images gives $\alpha=0.95\pm0.17$, but the scale radius $r_s$
and concentration $c$ are unrealistic.
(The models ensemble without the quint gives
$\alpha=0.5\pm0.5$, with the formal uncertainty increasing to
$\pm1.5$ at 99\% confidence.)
We find systematically different parameters depending on whether the
radius range with meso-structure is included in the fit: excluding the
meso-structure gives $\alpha=1.22\pm0.48$. We find smaller errors if
we put priors on the scale length $r_s$ \citep[as
in][]{2008arXiv0802.4292L}, but then the best fit simply chooses the
largest possible scale length.

It is easy to see why the derived $r_s$ is so unstable.  To
infer $r_s$ and $c$ from data interior to $r_s$, one would have to
accurately measure the gradual steepening of $\rho(r)$.  But in
Fig.~\ref{fig-surfden}, we see that in the well-constrained region,
$\rho(r)$ is probably getting shallower rather than steeper.  In other
words, in these data meso-structure is a stronger effect than scale
radius and concentration.

It is interesting to ask whether the above meso structure can be
explained simply by the dark matter associated with visible galaxies
in the cluster. There are five galaxies at a projected distance of
$\sim 100$\,kpc that could contribute. Four lie to the south east of
the cD galaxy; one to the north. We obtain a rather crude estimate of
the total mass in the meso-structure by subtracting the cumulative
mass of the best fit equation \ref{eq-nfwlike} for $r<100$\,kpc from
the cumulative mass of the ensemble average, and measuring the
residual over the range 100\,kpc$<r<$150\,kpc. This gives
$M_\mathrm{meso} = 7\times 10^{12}\,M_\odot$, which implies an average
mass per galaxy of $\sim 1.4\times 10^{12}\,M_\odot$. This indicates
that all the mass in the meso-structure could be associated with
visible galaxies \citep[as the model of]
[assumes]{2008arXiv0802.4292L} but does not require it.  Whether
associated with the visible galaxies or not, however, the
meso-structure must be a transient phenomenon. There are four cases of
interest. The meso-structure could be several galaxies or one large
group, or it could be viewed in projection, or really lie at $\sim
100$\,kpc. If projected along the line of sight, then it is transient
because it must rapidly move to larger projected radius. If it is really at
$\sim 100$\,kpc then, at the inferred densities of a few times
$10^6M_\odot\kpc^{-3}$, the crossing time is $\sim 0.2$\,Gyrs. If the
meso structure is comprised of individual galaxies, these will rapidly
phase mix; if instead the meso-structure is a large group then it will
be rapidly tidally stripped, since its density is significantly lower
than that of the main cluster \citep[see e.g.,][]{2006MNRAS.366..429R}.
In all cases, the meso-structure will be transient.

\subsection{Towards direct comparison of lenses and simulated clusters}

While an inferred inner profile of $\sim r^{-1}$ provides some
evidence in favor of universal profiles, it is desirable to be able to
compare lens reconstructions with theory and simulations more
directly.  Ideally, the comparison would involve quantities clearly
related to the cluster-formation process.  Failing that, one would
like to compare quantities that are well-constrained by the
observations.  Current parameterizations do not achieve either of
these.

With this in mind, let us consider the lensing deflection for a source
at infinity, as a function of projected radius
\begin{equation}
\ainf(R) = \frac{4GM(R)}{c^2R} .
\label{def-ainf}
\end{equation}
Here $M(R)$ is the mass within a circle of projected radius $R$.  The
deflection angle can also be expressed in velocity units: if we write
$\sigma^2(R)=c^2\ainf(R)/(4\pi)$ then for an isothermal sphere $\sigma$
will be the velocity dispersion.

While $\ainf(R)$ has no obvious connection to the formation process,
comparing `bending-angle curves' for galaxy and cluster simulations
(see Fig.~\ref{fig-sim}) indicates a clear qualitative difference: for
a galaxy $\ainf(R)$ tends to rise steeply at first and then stay flat
over an extended range, whereas for a cluster $\ainf(R)$ tends to rise
and then gradually turn over.  This is just another way of stating
that galaxies tend to have $\sim r^{-2}$ profiles and flat rotation
curves, whereas clusters tend to have shallower profiles and rising
rotation curves.  Put in still another way, galaxies have higher
concentration than clusters.

In lensing, $\ainf$ is the quantity best constrained.  Fig.~\ref{fig-ainf}
shows
$\ainf(R)$ for ACO~1703, reconstructed with and without the quint.
Also shown is what we may call the critical bending angle
\begin{equation}
\acrit(R) = \frac{\Ds}{\Dls} \frac{R}{\Dl} .
\label{def-acrit}
\end{equation}
An intersection
\begin{equation}
\ainf(R_E) = \acrit(R_E)
\label{def-RE}
\end{equation}
corresponds to an Einstein ring.  Strictly speaking, the Einstein
radius is only defined for a perfectly circular lens, but it is useful
to take (\ref{def-RE}) as a working definition of $R_E$.

From Fig.~\ref{fig-ainf} it is clear that all the sources except the
quint have a very similar $R_E$, whereas for the quint $R_E$ is
significantly less.  This tightly constrains $M(R)$ at two different
radii, and is the reason for the constraint on the profile.  Without
the redshift contrast provided by the quint, the profile becomes much
more uncertain, and even isothermal-type profiles are marginally
permitted.

It is interesting to note that the shallower the profile, the more
$R_E$ depends on the source redshift.  This can be inferred from
Fig.~\ref{fig-ainf}: for shallower profiles, $M(R)$ and hence
$\ainf(R)$ would rise more steeply, and hence the intersections with
the $\acrit$ lines would get more widely spaced.

In particular, we see $\ainf(R)$ curve and the $\acrit$ lines
intersecting at $R\simeq125\kpc$ for the high-redshift sources and at
$R\simeq75\kpc$ for the quint.  Had the $\ainf(R)$ curve been constant
at the $\ainf(R=125\kpc)$ value, the $\acrit$ line for the quint would
have intersected it at $R\simeq100\kpc$.  We may say that the
characteristic size of the quint is 60\% that of the other systems,
whereas for an isothermal lens we would expect it to be 80\% the size.
Thus the simple observation that the quad is much larger than the
quint is already an indicator that the profile is shallow.

More formally,
let us denote the ratio $\Dls/\Ds$ by $f$, and suppose that $M(R)$
mass varies in some range of $R$ as $R^\beta$.  Substituting in
Eqs.~(\ref{def-ainf}--\ref{def-RE}) it follows that
\begin{equation}
\frac{d\ln R_E}{d\ln f} = \frac1{2-\beta} .
\end{equation}
The point-mass case ($\beta=0$) has the weakest dependence on the
redshift contrast, while a constant-density sheet ($\beta=2$) becomes
infinitely sensitive.  An isothermal lens has $\beta=1$, and inner
clusters are expected to locally have $\beta>1$, and hence a small
change in $\Dls/Ds$ implies a larger change in $R_E$.

\section{Discussion}

This paper has been an analysis of the multiple-image lensing systems
reported by \cite{2008arXiv0802.4292L} in the cluster ACO~1703, and
comparison of the inner profile with what hierarchical structure
formation predicts.  The main conclusions are as follows.
\begin{enumerate}
\item It is easy to model the lens, and in simple qualitative features
  all models will agree.  But models are highly non-unique in
  important details.  Readers of the online version can explore models
  interactively: see Appendix \ref{online}.
\item Robust constraints can, however, be derived from ensembles of
  models.  We find the inner profile is well constrained and supports
  the prediction of universal profiles.  Similar work earlier on the
  clusters SDSS~J1004+411 and ACO~1689 led to the same conclusion
  \citep{2006ApJ...652L...5S}.
\item It is possible to identify where the inner-slope constraint
  comes from.  Behind the cluster at $z=0.28$ there is on the one hand
  a source at $z=0.888$ lensed into a quint, and on the other hand
  several sources at $z$ from 2.2 to 3.0 that are variously
  multiply-imaged.  The redshift contrast between the quint and the
  other sources is responsible for the inner-slope constraint. Without
  the quint, any constraint on the inner slope would have been much
  weaker, and even an isothermal-type profile would fit.
\item The projected-density contours (see Fig.~\ref{fig-mass})
  correlate with the cluster galaxies. Since the mass involved is much
  more than the stellar mass of the galaxies, this suggests
  meso-structure, i.e., extended dark-matter substructure correlated
  with galaxies.
\end{enumerate}

These results raise a further question: How can one most effectively
compare a reconstructed lensing cluster with simulations and/or
phenomenological models for structure?  A non-parametric test would be
very welcome.  We suggest that the deflection angle (and specifically
its redshift dependence) may form the basis for such a test.  In
particular, we note that the shallower the lensing profile, the more
sensitive the lensed images are to source-redshift contrast.

\acknowledgments We thank Andrea Macci\`o for providing the $N$-body
and hydro models used for comparisons and tests, John Stott for
providing a composite image of the cluster, Liliya Williams for
comments on the manuscript, and the referee, Dan Coe, for many useful
recommendations. 

\appendix
\section{Deprojection of the density profile}\label{ap-tests}

To derive the projected density $\rho(r)$ shown in Fig.~\ref{fig-mass}
we first computed the circularly-averaged $\Sigma(R)$ from the lensing
mass map and then applied an Abel deprojection
\begin{equation}
\rho(r) = -\frac{1}{\pi}\int_r^{\infty}
          \frac{d\Sigma(R)}{dR}\frac{dR}{\sqrt{R^2-r^2}} .
\end{equation}
To evaluate the numerical derivative and then the integral, we
linearly interpolate $\Sigma(R)$ up to the $R$ of the outermost image.
For the contribution beyond the outermost image, we assume $\Sigma
\propto R^{-2}$.  This makes the total mass formally divergent, but
that is harmless for the region of interest.  In fact, as noted in our
earlier work \citep{2006ApJ...652L...5S,2007ApJ...667..645R} and also
by \cite{2008arXiv0801.1875B}, the inferred $\rho(r)$ is very
insensitive to $\Sigma(R)$ for $R\gg r$, as long as $\Sigma$ is
asymptotically steeper than $R^{-1}$.

Since we have an ensemble of models, we automatically derive
uncertainties.  Previously, we deprojected the maximum and minimum of
an uncertainty band in $\Sigma(R)$ and showed the result as an
uncertainty band in $\rho(r)$.  This caused a problem in that the
upper and lower range of $\rho(r)$ could switch, giving the illusion
of zero uncertainty at certain $r$ \citep[see the right panels of
Fig.~1 in][]{2006ApJ...652L...5S}.  Here we deproject each $\Sigma(R)$
profile from the model ensemble separately, and then derive an
uncertainty band in $\rho(r)$.  At any $r$, the band represents the
uncertainty at that $r$, marginalized over all the other radial bins.

The above procedure assumes spherical symmetry and a possible concern
is that this may introduce a large systematic error if the cluster is
significantly non-spherical, as ACO~1703 evidently is.  To test for
this, we projected and then deprojected (using the above method) 14
triaxial cluster halos taken from a cosmological $N$-body simulation.
In Fig.~\ref{fig:test} we show the worst example --- where the cluster
is really two merging clusters.  We see that if the substructure is
evident in the projection, the spherically-averaged $\rho(r)$ is
mostly recovered to within the claimed uncertainties in the real
cluster.  The secondary cluster appears as a shelf $\rho(r)$ when it
is actually a bump.  The worst-case scenario is when the system is
projected `along the barrel' such that it appears spherical.  In this
case, the inferred $\rho(r)$ is overestimated by a about a factor of
two.

To summarize, the errors in the deprojection are generally smaller than
the uncertainty from lensing, except in the case where a major
substructure is hidden along the line of sight.

\section{Online modeling}\label{online}

\begin{figure}
\begin{center}
\epsscale{0.3}
\plotone{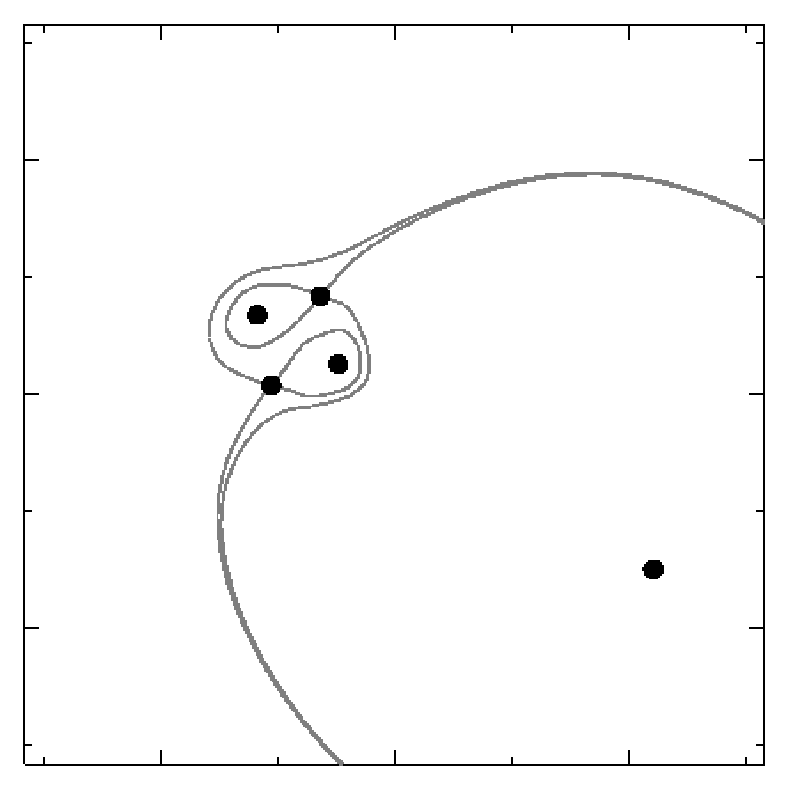}\plotone{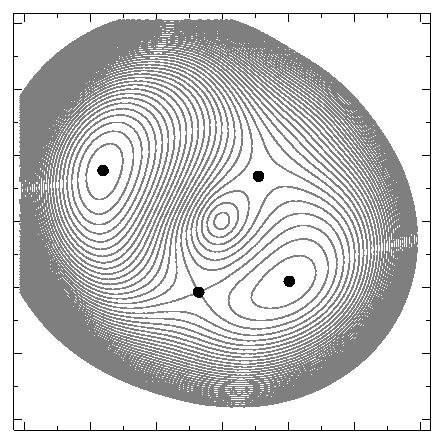}\plotone{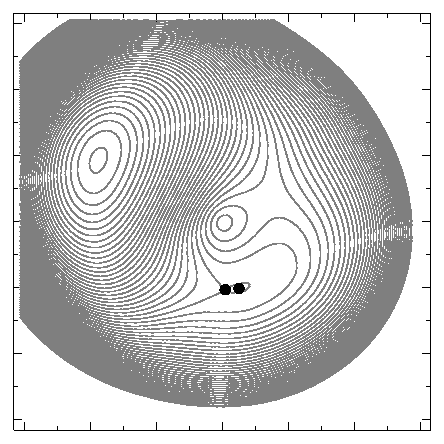}
\goodbreak
\plotone{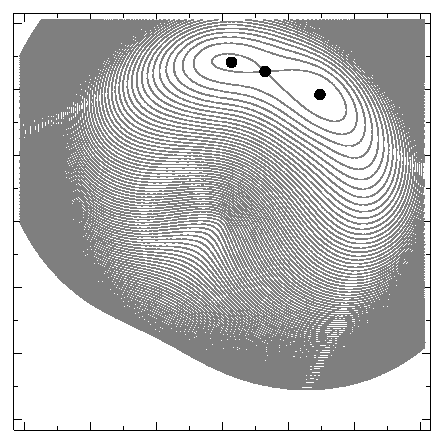}\plotone{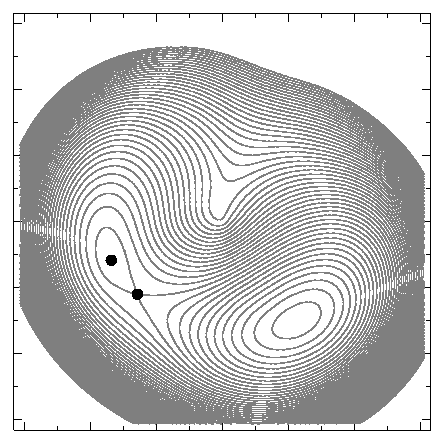}\plotone{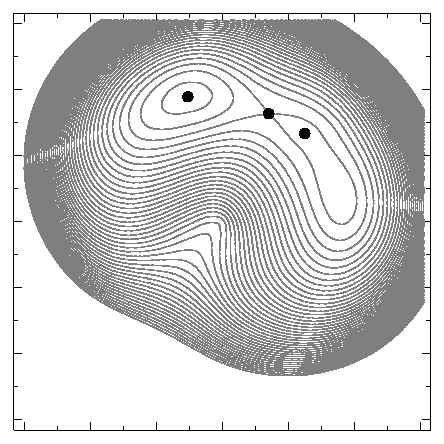}
\caption{Arrival-time contours for six of the multiple-image systems
  in the ensemble-average model (left panel of Fig.~\ref{fig-mass}).
  The panels, in reading order, refer to the quint \citep[1 in the
  numbering system of][]{2008arXiv0802.4292L}, a quad (16), the
  incipient quint (2), a northern lemniscate (7), a southern
  lemniscate (10), and a northern long arc (4). In all panels, ticks
  are $10''$ apart; thus the quint is shown zoomed in.  
  Each panel uses its own contour step, chosen to best illustrate the
  image configuration, but the step is always a few months of light
  travel time.}
\label{fig-arriv}
\end{center}
\end{figure}

\begin{figure}
\begin{center}
\epsscale{0.5}
\plotone{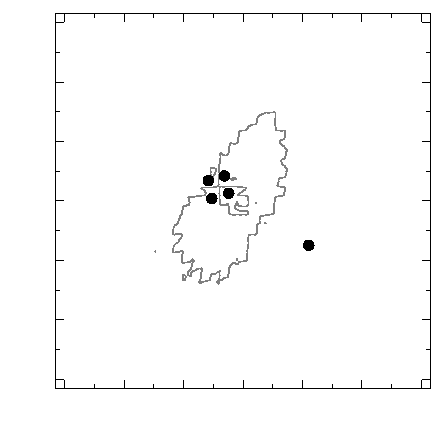}\plotone{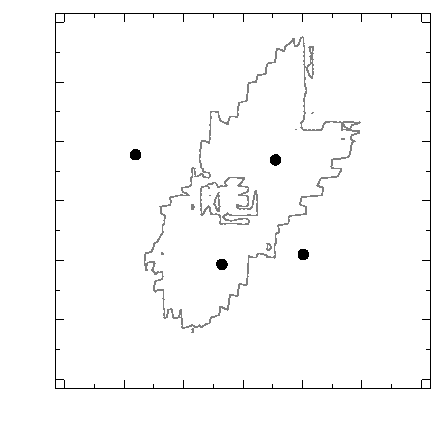}
\caption{Critical curves of the quint and a quad, in the ensemble
average model.  Qualitatively, these two panels resemble the middle
panels of Fig.~\ref{fig-simp}.  But note the size difference between
the quint and the quad.  Ticks are $10''$ apart.}
\label{fig-crit}
\end{center}
\end{figure}

\begin{figure}
\begin{center}
\epsscale{0.5}
\includegraphics[width=0.49\textwidth]{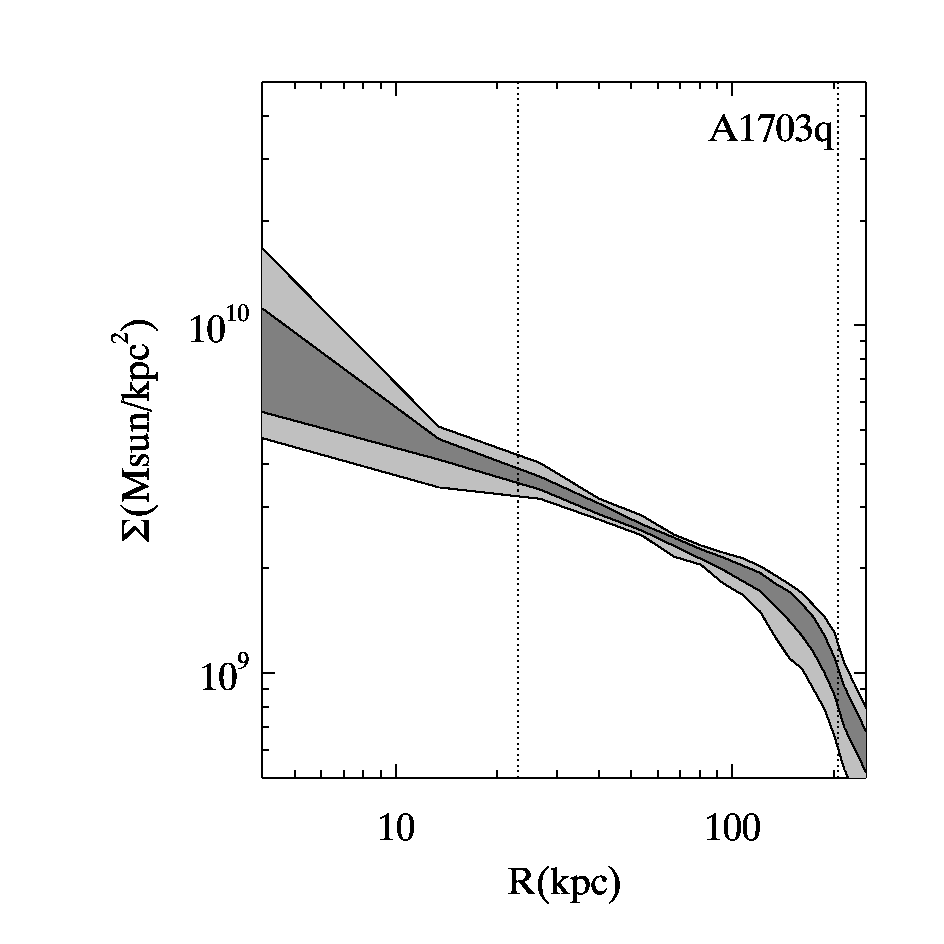}
\includegraphics[width=0.49\textwidth]{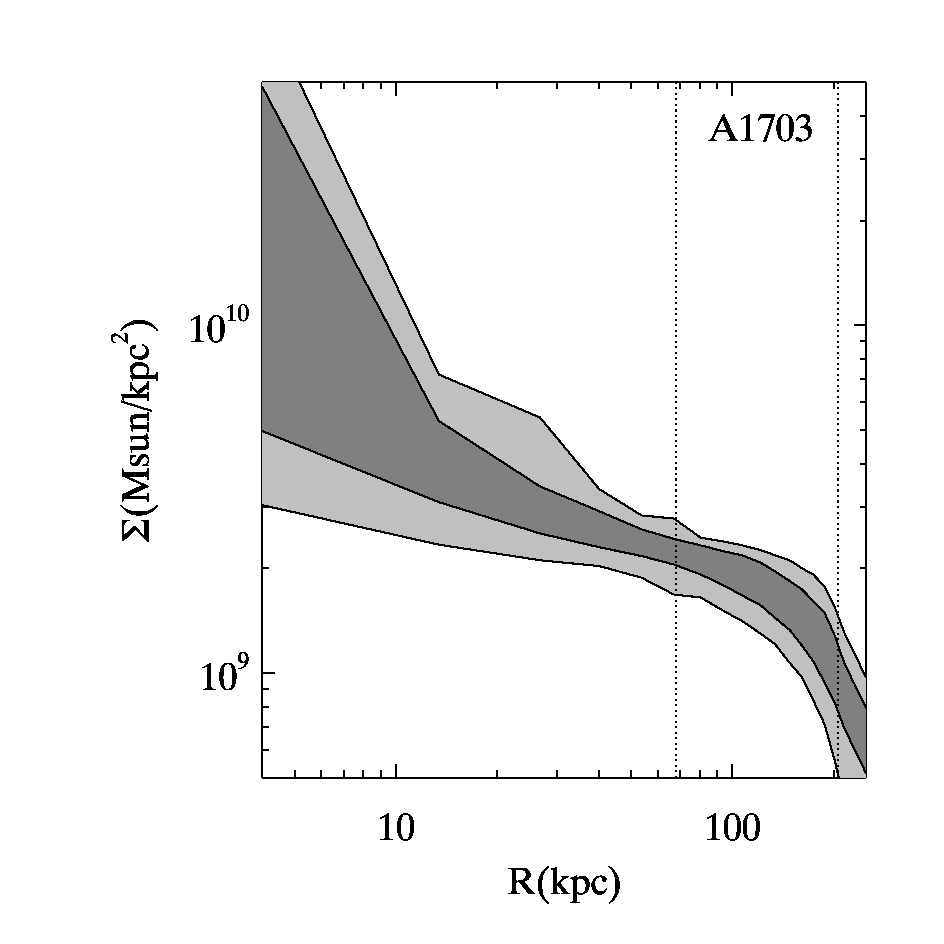}\\
\includegraphics[width=0.49\textwidth]{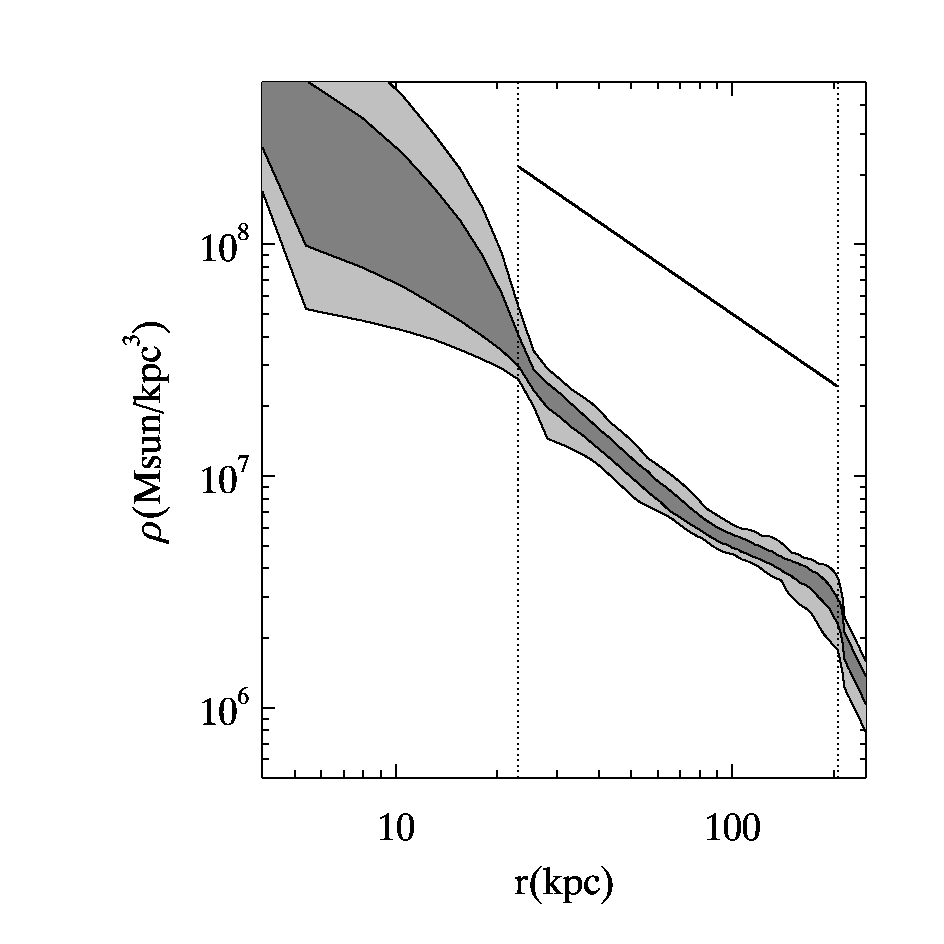}
\includegraphics[width=0.49\textwidth]{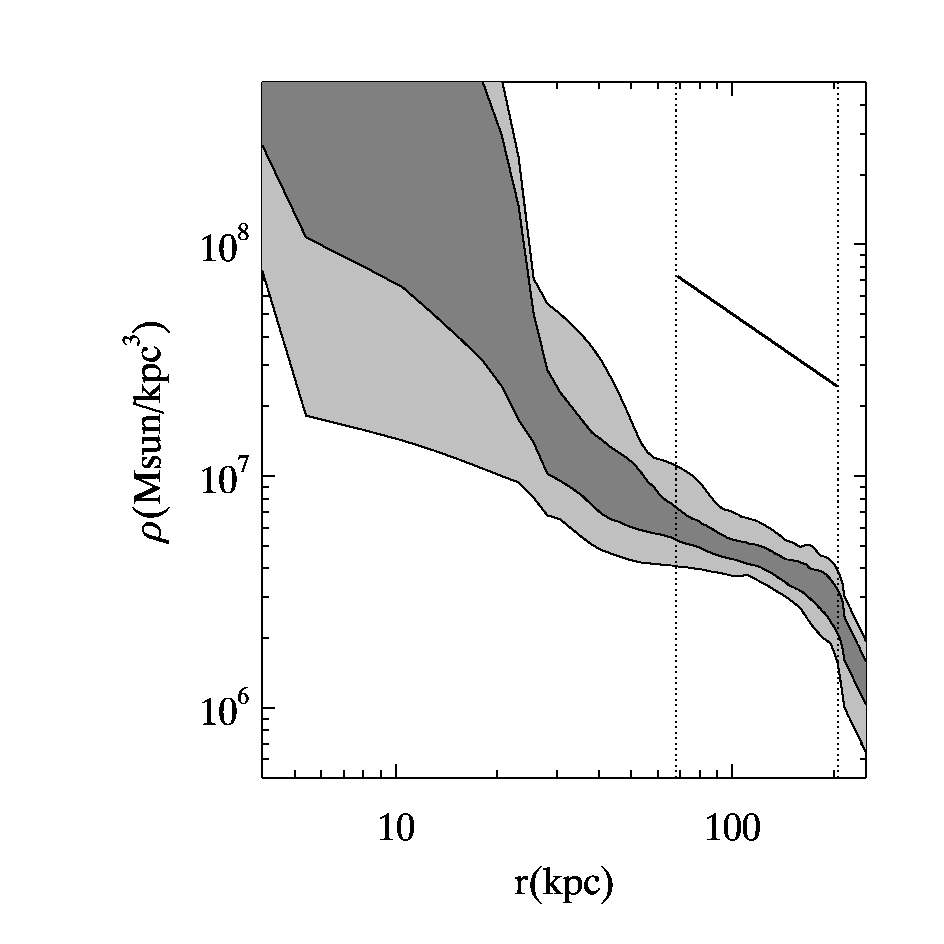}\\
\caption{Surface density $\Sigma(R)$ (top panels) and deprojected
  density $\rho(r)$ (bottom panels) for ACO~1703.  The left column is
  for models with the quint, the right column for models without the
  quint. The dark gray and light gray bands give 68\% and 99\%
  confidence intervals respectively. The vertical dotted lines are the
  projected radii of the inner and outermost image.  The oblique
  line in the bottom panels shows $r^{-1}$.  With the quint
  included, the inner profile is constrained to be quite close to
  $\rho\propto r^{-1}$. Without the quint, even $\rho\propto r^{-2}$
  is admissible.  Notice also the `shelf' in $\rho(r)$ for $r\sim
  100\kpc$, which may indicate meso-structure (see text).}
\label{fig-surfden}
\end{center}
\end{figure}

\begin{figure}
\begin{center}
\epsscale{0.3}
\plotone{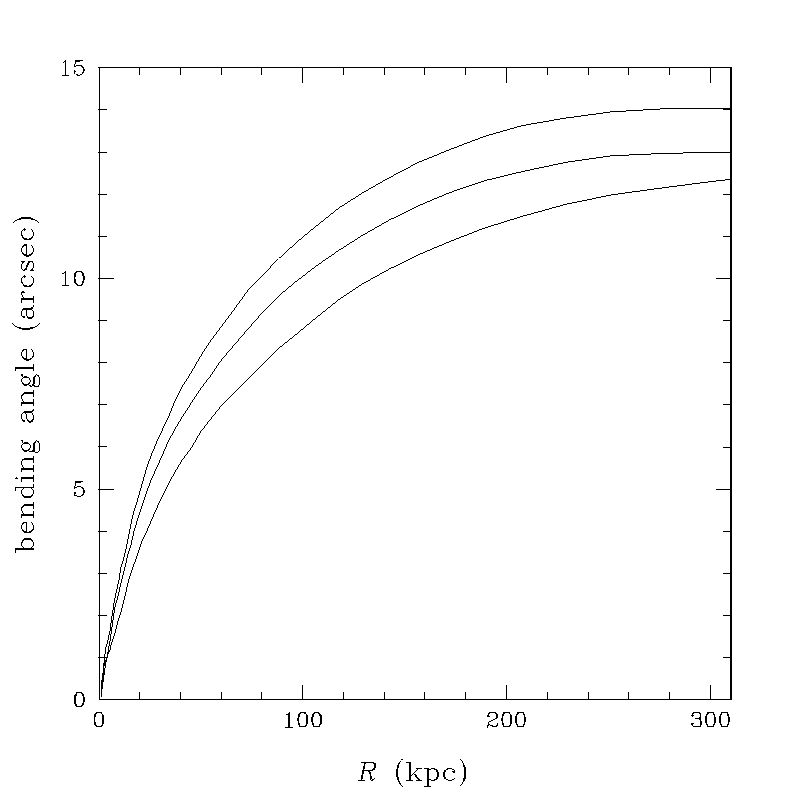}
\plotone{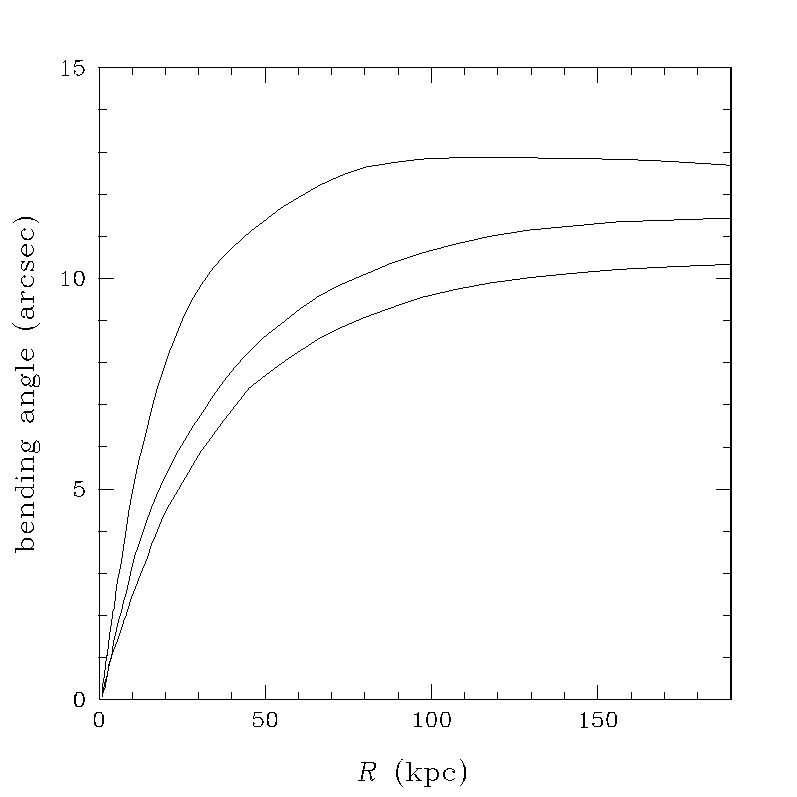}
\plotone{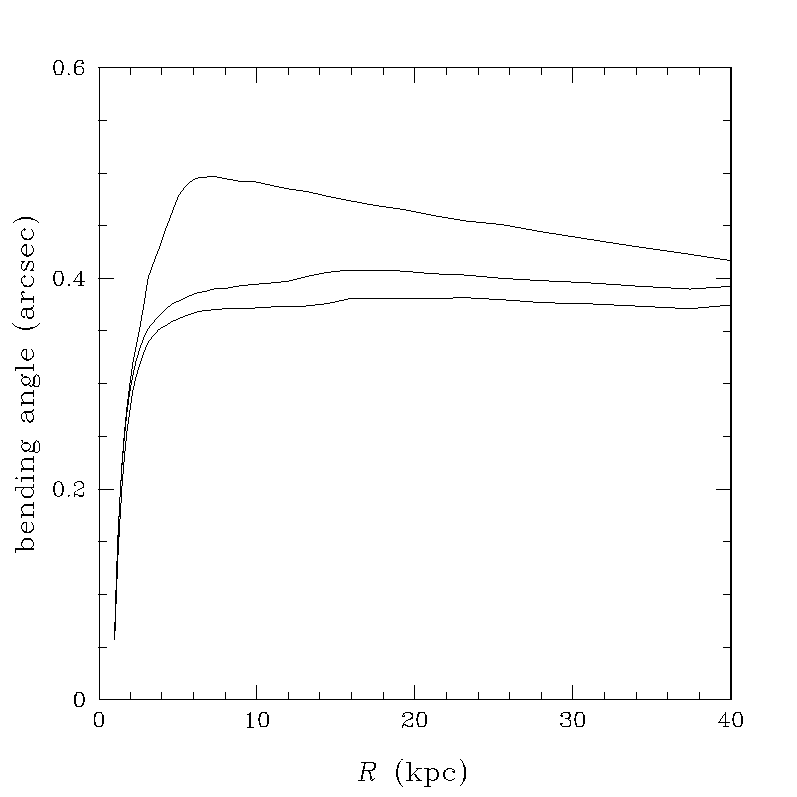}
\caption{The bending angle $\ainf(R)$ for two simulated clusters (left
and middle panels) and a galaxy (right panel), out to $\sim25\%$ of
the total mass.  The three curves denote three orthogonal projections
of a simulation to produce a lens.}
\label{fig-sim}
\end{center}
\end{figure}

\begin{figure}
\begin{center}
\epsscale{0.6}
\plotone{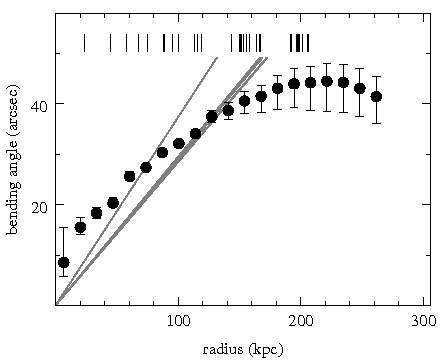}
\plotone{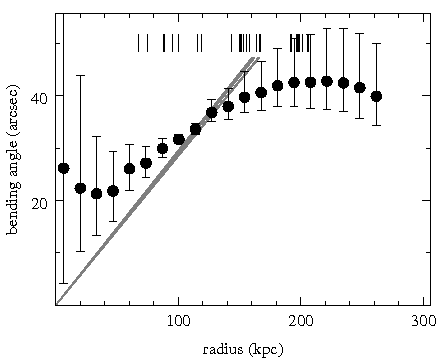}
\caption{Deflection angle $\ainf(R)$ of the circularly averaged
profile (for sources at infinity) against projected radius $R$.
The upper panel come from a
models with the quint included, and the lower panel from models with
the quint excluded.  The points show $\ainf(R)$, with 90\%
uncertainties derived from the ensemble, against the projected radius
$R$. The barcode-like patterns show the $R$ values of individual
images.  The oblique lines show the critical deflection $\acrit(R)$
for each source redshift; note how the quint stands out.}
\label{fig-ainf}
\end{center}
\end{figure}

\begin{figure}
\begin{center}

\includegraphics[width=0.3\textwidth]{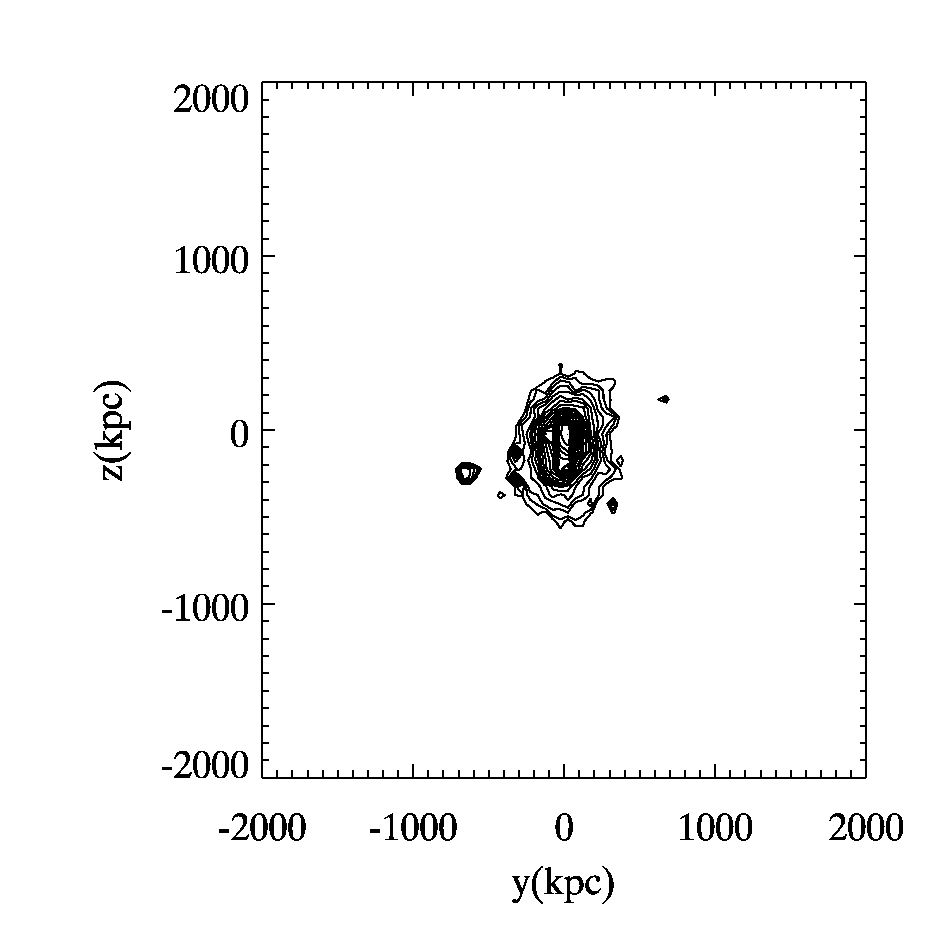}
\includegraphics[width=0.3\textwidth]{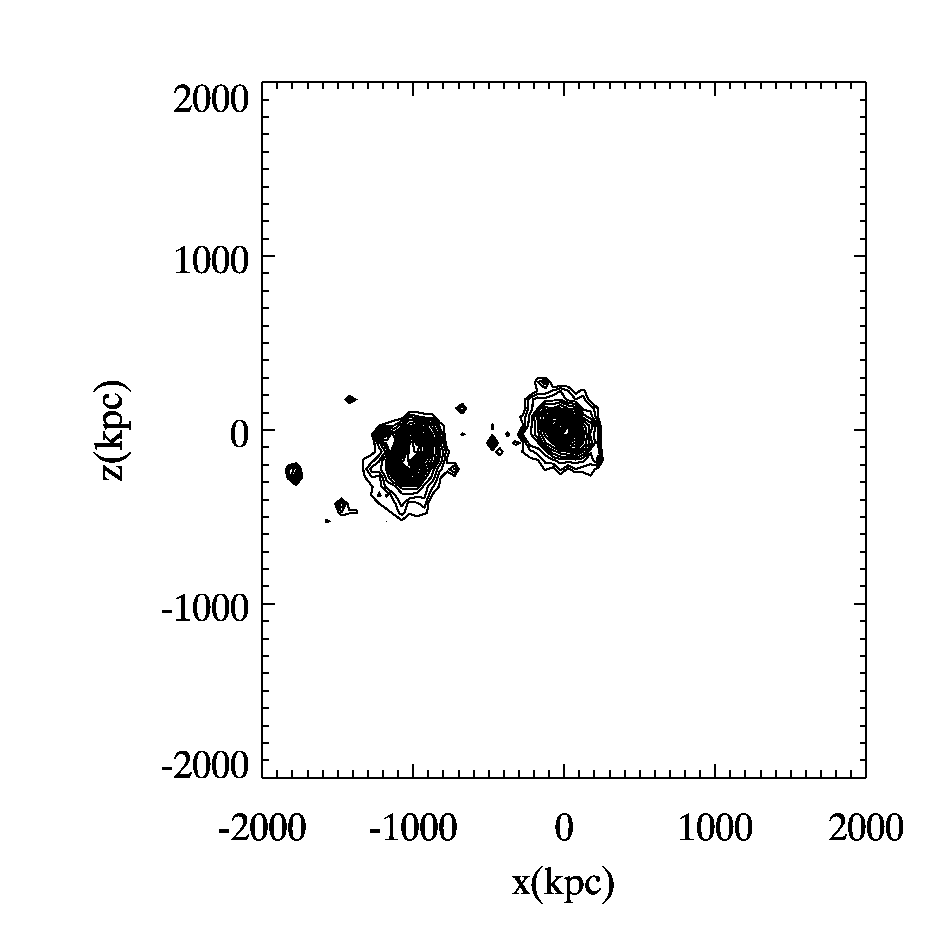}
\includegraphics[width=0.3\textwidth]{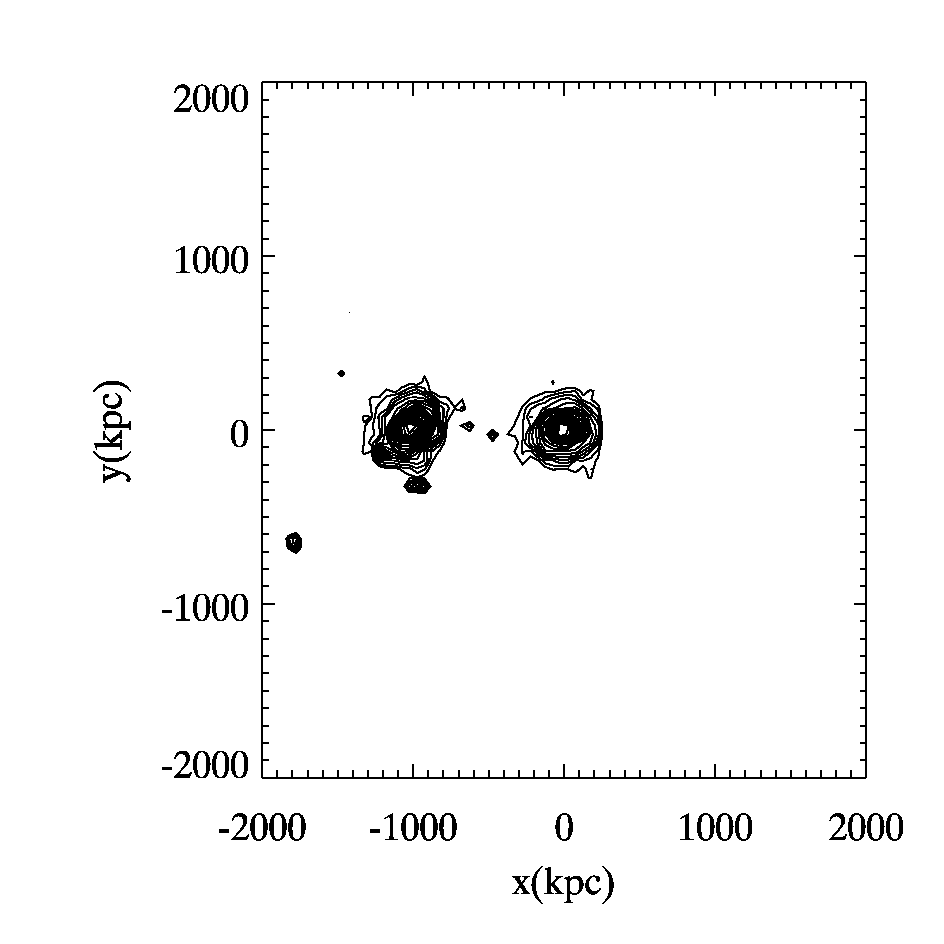}\\
\includegraphics[width=0.3\textwidth]{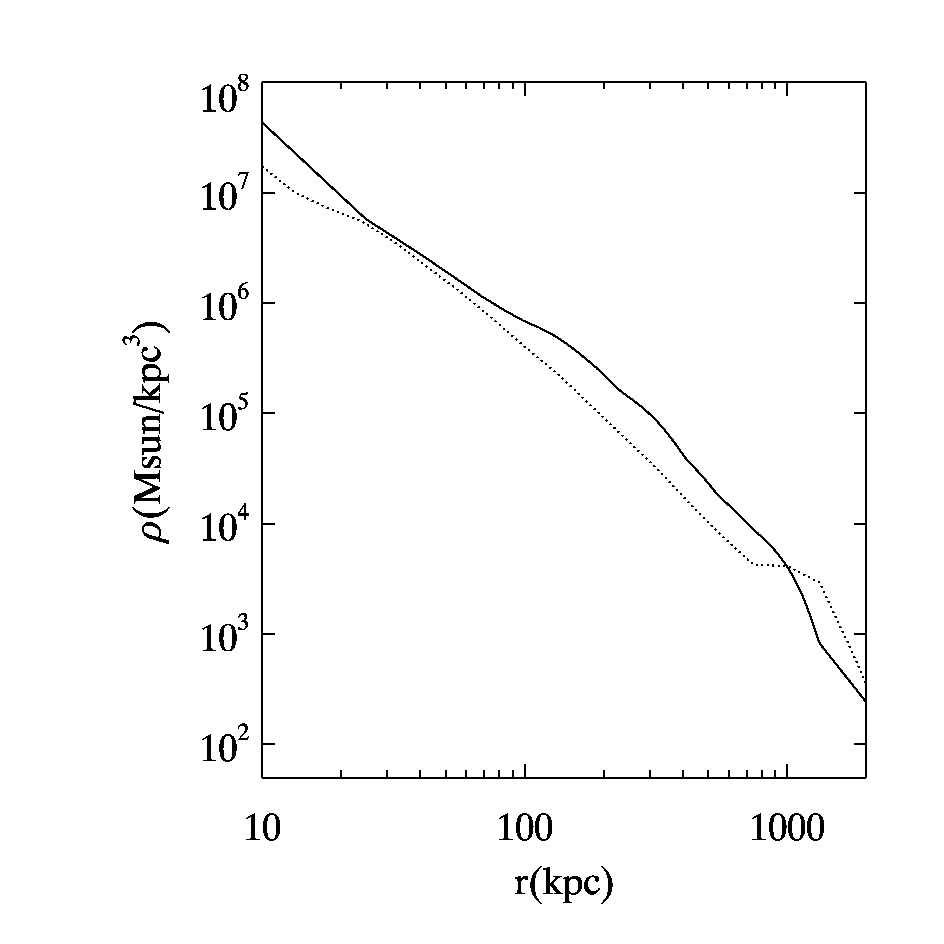}
\includegraphics[width=0.3\textwidth]{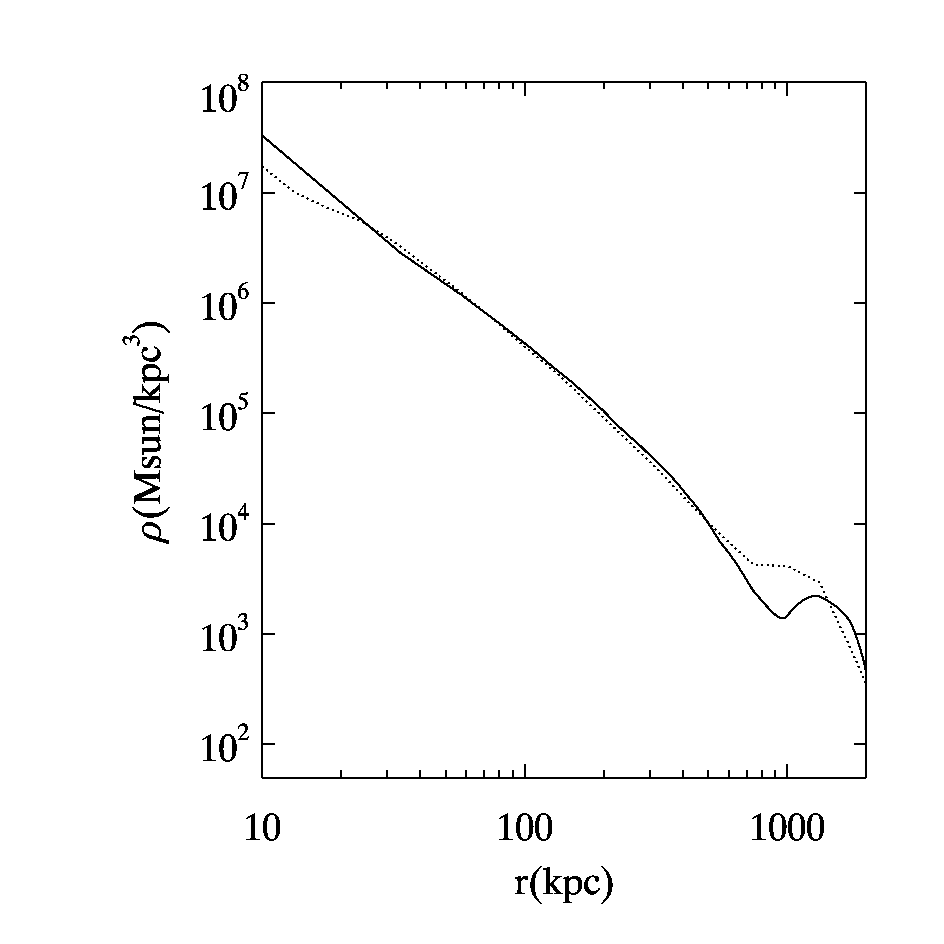}
\includegraphics[width=0.3\textwidth]{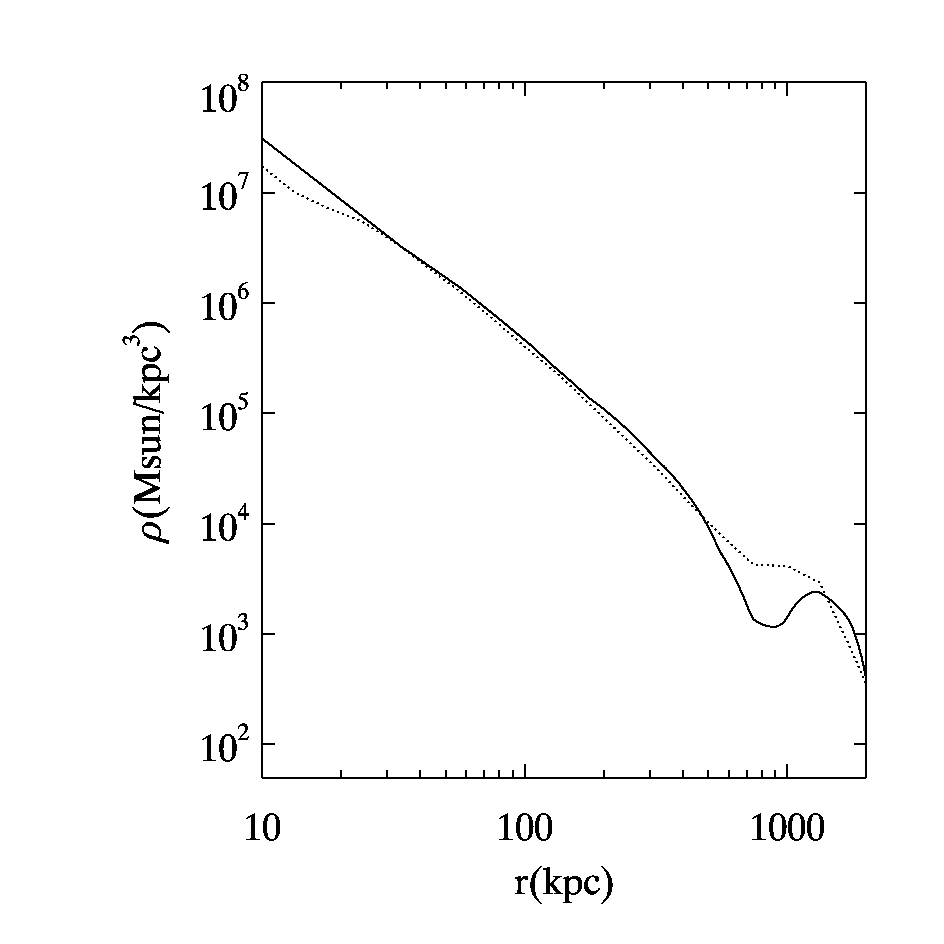}\\
\includegraphics[width=0.3\textwidth]{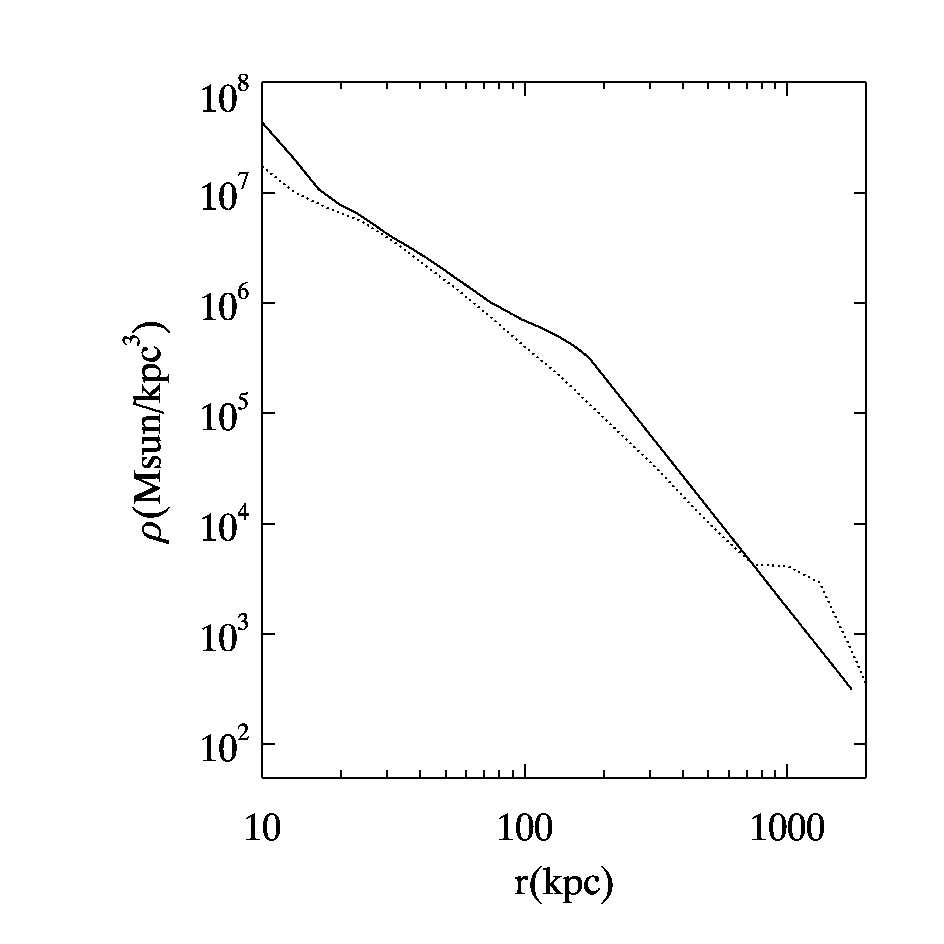}
\includegraphics[width=0.3\textwidth]{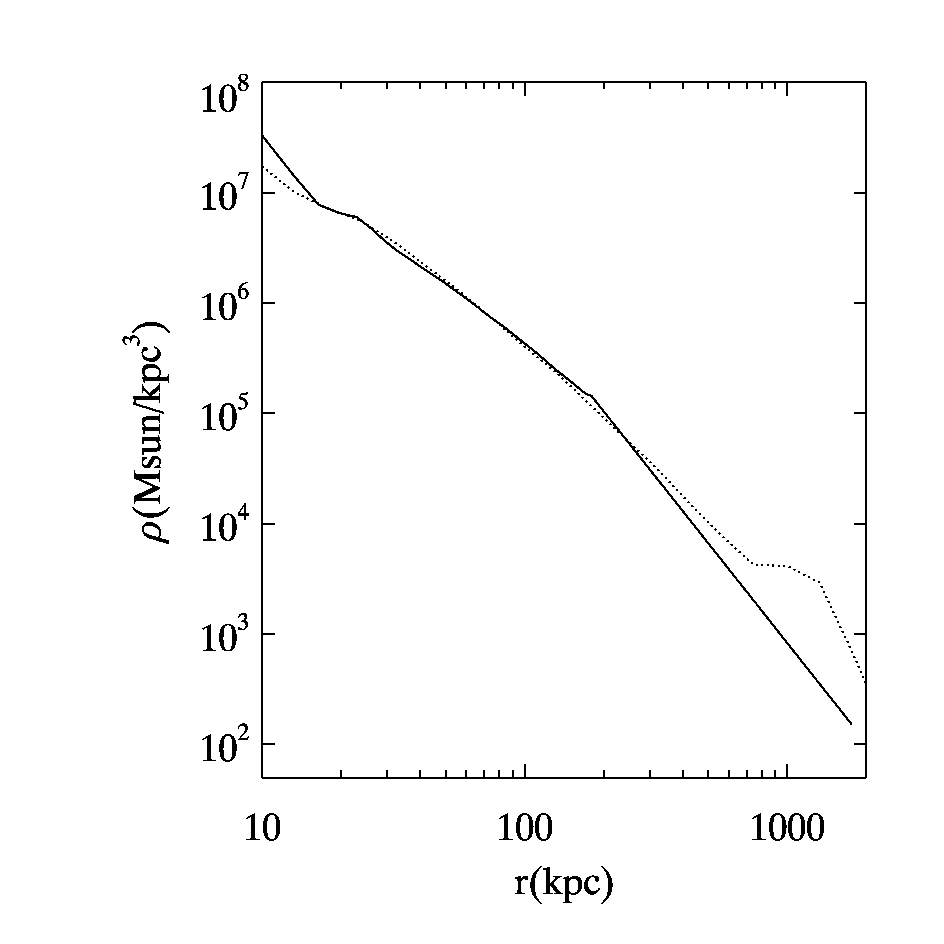}
\includegraphics[width=0.3\textwidth]{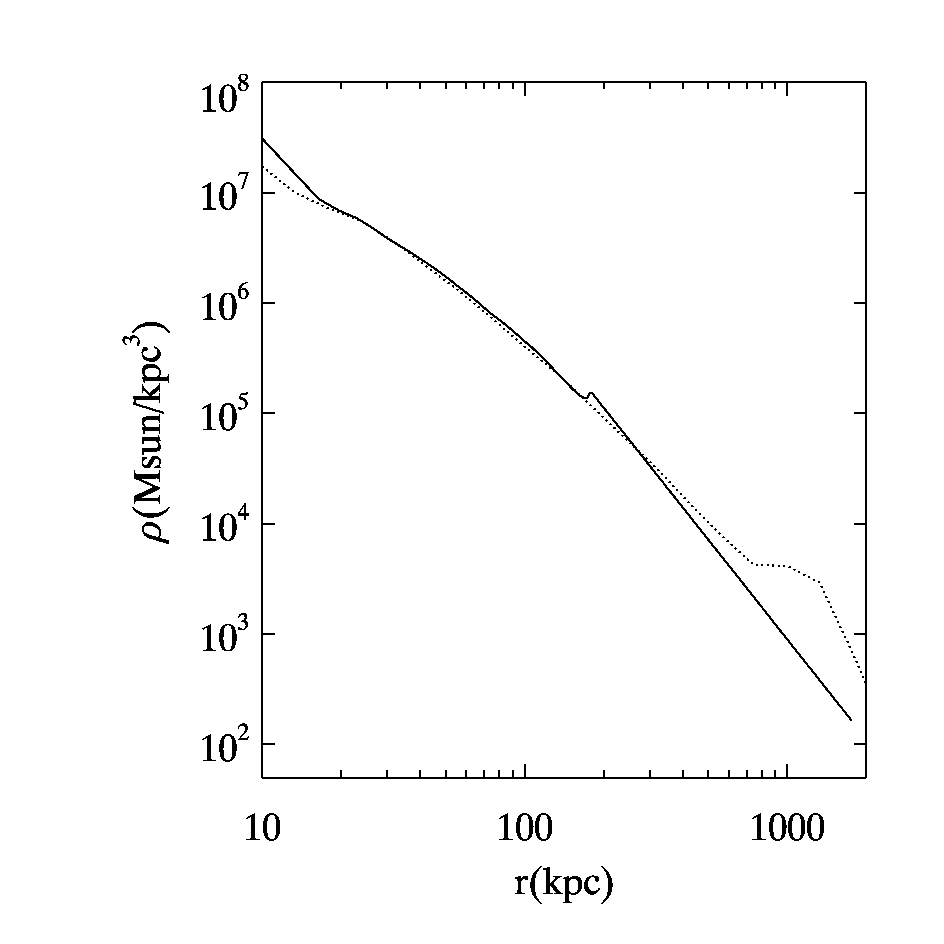}

\caption{Test of the Abel deprojection method for an extremely
  non-spherical system.  The upper panels show the density of an
  $N$-body system of two merging clusters projected along three
  orthogonal axes.  The middle panels show the density $\rho(r)$
  spherically averaged with respect to the centroid of the more massive
  cluster (dotted curves), and the value recovered by taking the
  circularly averaged projected density $\Sigma(R)$ and then
  deprojecting (solid curves). The lower panels are similar, except
  that $\Sigma(R)$ has been truncated at $200\kpc$ to mimic the strong
  lensing regime.  The worst case is in the left column, when the two
  clusters are aligned along the line of sight.}

\label{fig:test}
\end{center}
\end{figure}

The online version of this paper includes the {\em PixeLens\/}
modeling program as a Java applet.  Readers can model the lens
interactively within a web browser.  The example input uses lower
resolution than the paper, and a subset of the image systems (in fact,
the six systems illustrated in Figure~\ref{fig-arriv}), but still
enables one to verify the significance of redshift contrast.  The
input can be edited online, and additional data typed or pasted in.

The input syntax is as described in \cite{2004AJ....127.2604S} but
with one important new feature, namely multiple source redshifts.  An
$M$-image lens system with source redshift $z_S$ is input as
$$ \matrix{ \hbox{\tt multi} & M   & z_S    \cr
                         x_1 & y_1 & p_1  \cr
                             & \ldots \cr
                         x_M & y_M & p_M  }
$$
where $(x_i,y_i)$ are the image positions and $p_i$ encode the image
types (1 for a mininum, 2 for a saddle point, 3 for a maximum).
\cite{2008ApJ...679...17C} describes recent developments of the
modeling technique, including parallelism.

\bibliographystyle{apj}
\bibliography{ms.bbl}

\clearpage

\end{document}